GIGSAATHI

by the internet Research Lab

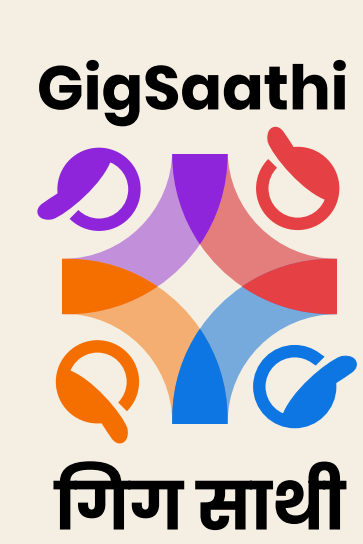

# Forced Volatility: Earnings and Incentives for Gig Work in Quick Commerce

**Research Team:**

Abhishek Sekharan
Ameya Kasliwal
Ambika Tandon
Gurshabad Grover
Javed Siddiqui
Omir Kumar

**Software Development:**

Kiwimesh

**Design:**

Srishti Verma

internet
Research
Lab

This project was partly funded by the Association for Progressive Communications' FeministTechJoy program

# Introduction

## Platform power & precarious pay

Quick commerce platforms deliver consumer goods within minutes. They have expanded rapidly and changed logistics supply chains globally. Instead of relying on traditional centralized warehouses, these platforms utilize a dense network of neighborhood-level "dark stores" – strategically located micro-fulfillment centers that typically stock over 2000 different products. Pioneered by firms like Getir and Glovo, this model markets convenience to customers through hyperfast doorstep delivery of groceries and daily essentials.

In India, this sector is experiencing explosive growth, far outpacing Amazon-style ecommerce. While the latter grew at 14% annually, quick commerce expanded at 73% in FY 2023-24, driven by companies like Zepto, Swiggy Instamart, and Blinkit (Sharma, 2024). Valued at $3.34 billion in 2024, the quick commerce industry is projected to reach $9.95 billion by 2029.

Despite being present in over 100 cities, quick commerce platforms are heavily concentrated in major metros, with non-metro cities contributing just over 20% of market value – a stark gap compared to their 60-70% share of overall retail.

Against this background, the sector is a catalyst for India's gig economy, with a projected

60% surge in gig worker hiring in 2025. This reliance on a vast, flexible delivery workforce poses challenges related to wages, working conditions, and social security.

A core feature of digital platforms is that they act as double or triple sided marketplaces, levying an entry or onboarding fee on workers to access work, while also selling services to customers for a price (Bricka and Schroeder, 2022). These platforms positions themselves as service providers for everyone who transacts through them: consumers, workers, sellers (for ecommerce), restaurants (for food delivery), etc.

They use algorithms to determine pay, allocate tasks, review performance, among other aspects (Jarrahi and Sutherland, 2019; Chan, 2022). Task allocation and wages change according to variable factors aimed at optimising profits.

A large proportion of the pay is contingent on daily, weekly, or monthly order-based incentives, which push workers to log longer hours, and in times and locations where there is an increase in demand. Gig workers are nudged to follow behavioural patterns through pay structures and in-app notifications — a process called gamification (Krzywdzinski and Gerber, 2021).

Companies have fed into the insidious narrative that algorithms are neutral and objective. Algorithmic 'black boxes' control key aspects of work, to the extent that even workers' immediate managers or supervisors may not fully understand or control their systems. Highly volatile algorithmic systems create an information asymmetry, leaving workers to grapple with variable task allocation and earnings.

Workers try to develop strategies to maximise pay but don't have accurate knowledge of how these systems work. Data shown to riders about future work is rare and sparse: Blinkit and Instamart do not present advance rate cards or incentives for upcoming days and weeks.

Lack of access to past earnings data is also a critical issue. Past data is systematically erased. Blinkit, Instamart and BigBasket show six months, one month and only 2 weeks of past data respectively, with no details on per order earnings. Unavailability of earnings data makes financial planning, control, and other work-related decisions.

Opaque wages weaken workers'

***Highly volatile algorithmic systems create an information asymmetry, leaving workers to grapple with variable task allocation and earnings. Workers try to develop strategies to maximise pay but don't have accurate knowledge of how these systems work.***

***Hiding and deleting past payment data curtails data transparency, making it difficult to calculate average earnings and even preventing workers from engaging in coordinated organising around wages.***

collective bargaining power against fluctuation, delays and deductions. Hiding and deleting past payment data curtails data transparency, making it difficult to calculate average earnings and even preventing workers from engaging in coordinated organising around wages.

Information asymmetry makes it almost impossible for gig workers to collect and present evidence regarding earnings and fluctuations in payouts over time and place, severely hindering legal and policy advocacy.

This project was aimed at addressing a key dynamic of the power relationship between digital platforms and gig workers: the information asymmetry on wages. It was implemented in partnership with the Rajdhani App Workers Union (RAWU), a union of gig workers in Delhi National Capital Region, associated with the Centre of Indian Trade Unions (a central trade union organising workers since 1970).

Our project piloted a chatbot allowing delivery workers in quick commerce to systematically collect their earnings. The chatbot would then process these screenshots to produce time series data on earnings, which we could then be used by workers to gain better insight into their own wages.

Our analysis reveals that there is significant variability in earnings in wages (base pay and incentives), working hours, and distance travelled. A significant proportion of wages are contingent on invisible or opaque variables that are not reported to workers.

Even the biggest company, Blinkit, does not have a minimum base pay per kilometer. Rate cards change multiple times daily through factors such as order size, weather and number of riders available, with shift-based incentives, long distance or rain bonuses, and penalties.

Wage setting is opaque and complex, calculated by a mix of algorithms and some manual controls by managers (during rain or strikes). In some companies such as BigBasket, setting and disbursing workers' wages are outsourced to third party vendors.

We also establish little to no correlation between earnings and time spent per order, which suggests that workers are not compensated proportionately for the total time spent in delivering a given order. We also find that incentives make up about a

*Our analysis reveals that there is significant variability in earnings in wages (base pay and incentives), working hours, and distance travelled. A significant proportion of wages are contingent on invisible or opaque variables that are not reported to workers.*

quarter of weekly earnings on average, again with very high variability.

Finally, we statistically demonstrate how quick commerce platforms deploy incentives to control when, how much, and where someone works. We found an inverse statistically significant relationship between earnings per order and incentives, which means that higher earnings per order have lower incentives and vice versa.

Our findings confirm that platforms promote dependence on incentives and gamified features of control, aiming to bind workers to the platform.

*We also find that incentives make up about a quarter of weekly earnings on average, again with very high variability.*

# Methodology

## Developing GigSaathi and onboarding workers

We conducted a brief needs assessment at the beginning of our pilot, to determine the extent of wage fluctuation and information opacity in four sectors within delivery work: ecommerce, food delivery, quick commerce, and bike taxi. We found that quick commerce and food delivery have very high levels of fluctuation, with the former being more opaque.

Opacity flows directly from the fragmented nature of the business model in quick commerce, which relies on a dense network of small warehouses in consumer neighbourhoods. We therefore narrowed down on quick commerce for the pilot.

We chose to analyse one of the key markets for quick commerce, New Delhi, focusing on the dominant player in the city, Blinkit. We selected a single platform to ensure uniformity in format and types of data. We selected three regions within Delhi and two warehouses in each. The core aim of the data collection was to track individual workers' earnings over time at each store.

At the beginning of the pilot we selected 5 Blinkit dark stores in Delhi. The 5 stores were located in the areas of Sangam Vihar (2), Green Park (2), and Zamrudpur (1). Most dark stores were set up in small and cramped basements or godowns, where basic facilities like washrooms and resting spots were lacking (RAWU, 2025). Some stores even lacked proper parking for bikes, and access to clean water. As the pilot progressed, we reduced the number of stores to 3 (Green Park and Zamrudpur) due to the non-responsiveness of workers.

We set up a WhatsApp chatbot that onboarded gig workers to GigSaathi, our data collection and analysis platform. Workers could share screenshots of their weekly payout information on our chatbot. In the beginning of the pilot we relied on online communication and physical visits to remind workers to submit their data. Over time, we increased our in-person visits as workers did not respond to online communication (reminders on WhatsApp from our team and through the chatbot). Overall, we conducted 8 in-person visits.

The initial strategy to collect weekly earnings screenshots from workers over a period of time (and then provide them information about their earnings)

Figures 1 and 2: Onboarding flow on GigSaathi

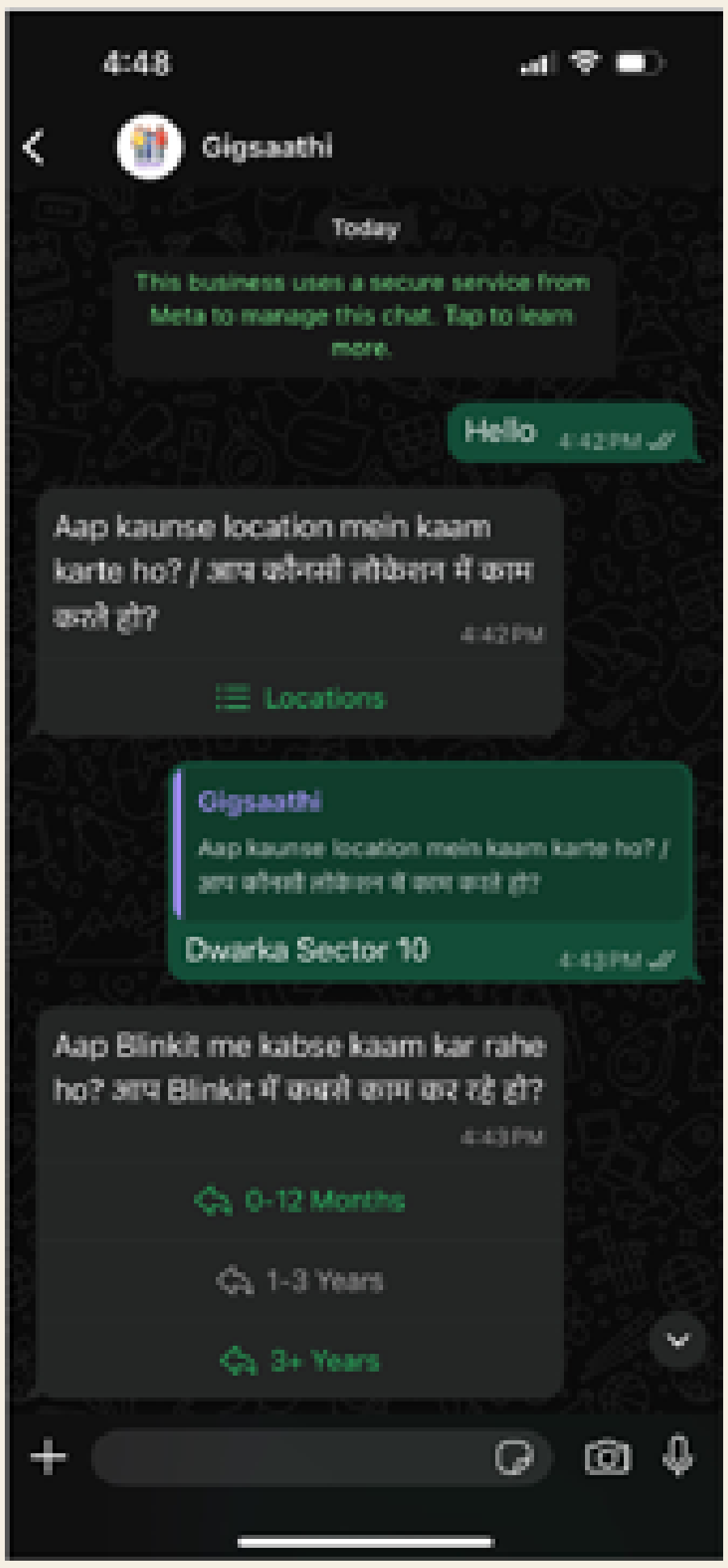

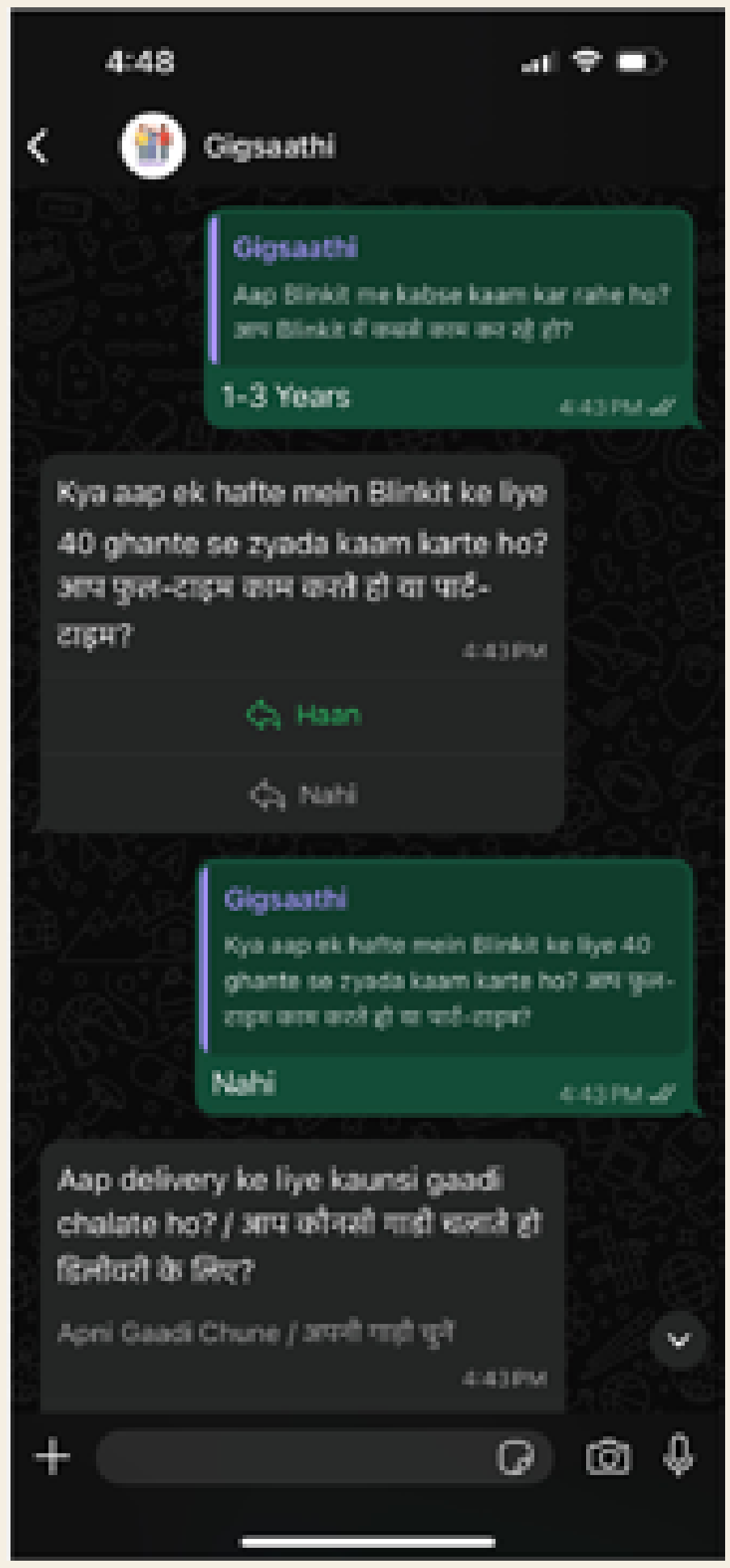

did not pan out as planned, as onboarded workers did not submit the data to the chatbot regularly.  We then pivoted to collecting daily incentive data to understand the logic and parameters deployed by platforms to set incentives.

**Developing GigSaathi**

The technical tool we developed was ‘GigSaathi’, a Whatsapp bot that automates collection and enumeration of earnings data from gig workers.

When a person first contacts GigSaathi, the bot prompts the person to share the (1) Blinkit location they work out of; (2) whether they work more than 40 hours per week on average for Blinkit; and (3) the type of vehicle they use and whether they own it.

Based on our initial on-ground interviews, we made all messages and bot interactions available in Hindi (in both Devanagari and Latin scripts). The GigSaathi bot collected weekly earnings/payout data in the form of screenshots (Figures 1 and 2).

GigSaathi was developed iteratively. In the beginning the bot prompted onboarded workers once a week on Monday to share their earnings data for the previous week to coincide with when Blinkit shared this data with workers.

We faced low response rates through this strategy, and programmed the bot to send daily reminders to those who had not shared their data, including nudges that highlighted the importance of worker-led data collection.

GigSaathi’s backend analysed the screenshots and parsed and enumerated the earnings data into a database that was accessed by our researchers.

**Onboarding and data collection :process and challenges**

We set a target of onboarding 10-12 workers from each store to send consistent data on the chatbot. During initial visits, we approached individual workers and talked to them about the payouts at the store. Workers often responded with frustration over the lack of transparency of wages.

For instance, a worker in Zamrudpur said they often have to travel longer distances than shown on the app for which they are not paid. Workers often went on strike to vent their frustration, but many in Green Park bemoaned how managers increase payouts for a few days after such action and then lower wages even below what they initially organised against. We introduced the chatbot as a tool that can help them gather proof for these fluctuations and even present it to the the company. We found that workers agreed with the core premise and aims of this exercise.

We tried two methods while approaching workers to talk about the GigSaathi chatbot. The first pitch involved talking about the general issues in the gig economy like absence of minimum wages, regulated working hours or social security, and the need for collective action. The second pitch started with talking about lack of transparency in earnings and working hours data. We found that the first pitch resonated more with workers.

We shared with workers the benefits of submitting their earnings related data on the GigSaathi chatbot, which include (1) analysis of data on earnings and working hours, (2) comparison of earnings with minimum wages, (3) the opportunity to collate data on expenses, such as internet connection, fuel, and bike rentals. However, such rich analysis was only possible after we had sufficient data to work with. The lack of immediate value dissuaded workers from joining GigSaathi.

For workers that remained interested, we provided them with help on getting started with the chatbot. Given the high level of digital literacy needed for gig work on online platforms, we expected our initial chatbot design to work smoothly, but workers often faced issues with dropdown options and other aspects of bot onboarding. We iterated the chatbot design to resolve these issues.

The biggest challenge was sustaining long-term interest in the project, despite adopting multiple ways to sustain data collection. Workers also dropped out of store-level WhatsApp groups we created, and some did not respond to calls or texts – highlighting the difficulty of establishing trust and interest in engagement with automated systems.

A larger concern in implementing any intervention in this workforce was its very high rate of attrition. A majority of the workers were in their early or late 20s, and migrants to Delhi from nearby states like Bihar, Uttar Pradesh, and Haryana in search of jobs; and were overwhelmingly men. Some of them worked for very short periods in the sector, and were often either local workers who work part-time for supplementary income or full-time seasonal migrants who travel frequently between the city and their native villages. Maintaining consistency in contact was difficult, and in each field visit our team encountered new workers at the same warehouse. During our in-person visits to dark stores for reminders, we were unable to locate many of the workers who had originally been onboarded. Some had already left the platform, while others were working different shift timings.

Given the challenges outlined above, we had to alter the scope of the project and our strategies. We reduced the number of dark stores to 3, and focused on more regular visits to engage workers who are associated with a dark store for longer periods.

The challenges also changed the scope of our data analysis. We initially planned to rely on consistent data on weekly earnings from the same cohort of workers to create a time series dataset on wages. However, given the challenges to collect such information, we also started collecting data on daily incentives and individual orders from any worker we could come across, so that these could be analyzed and yield useful insights on wage fluctuations and components. The following section details out analysis of payout data from individual orders and daily incentives.

# Findings

## More work ≠ more pay

***I. Spending more time on an order does not always translate to higher earnings. Delivery workers on quick commerce platforms remain uncompensated for a whole range of labour they do everyday***

We analyzed payout data from 41 orders reported between June-December 2025. The dataset consisted of: distance travelled per order (in kilometers), time spent per order (in minutes) which excludes waiting time in between orders, and earnings per order.

We started by cleaning the raw data, which included identifying outliers and missing data points. Based on the cleaned data we imputed two variables: earnings per km to analyze if longer distances translate to higher wages, and earnings per minute to analyze if longer order time corresponds to higher compensation.

Preliminary analysis of descriptive statistics reveal that on average, workers earn approximately Rs. 48 ($0.53), travel around 5 kms., and spend approximately 20 mins per order. Workers on average make approximately Rs. 12.18 ($0.13) per kilometer travelled and Rs. 2.9 ($0.03) per minute of order time.

However, our data indicates very high variations in each of these as expressed by high standard deviations (Rs. 23.5 per order, 2.74 kms per order, and 11 mins per order respectively). Earnings per kilometer had a standard deviation of 7.32, again indicating

Figure 3: Scatterplot of distance travelled per order and earnings per order

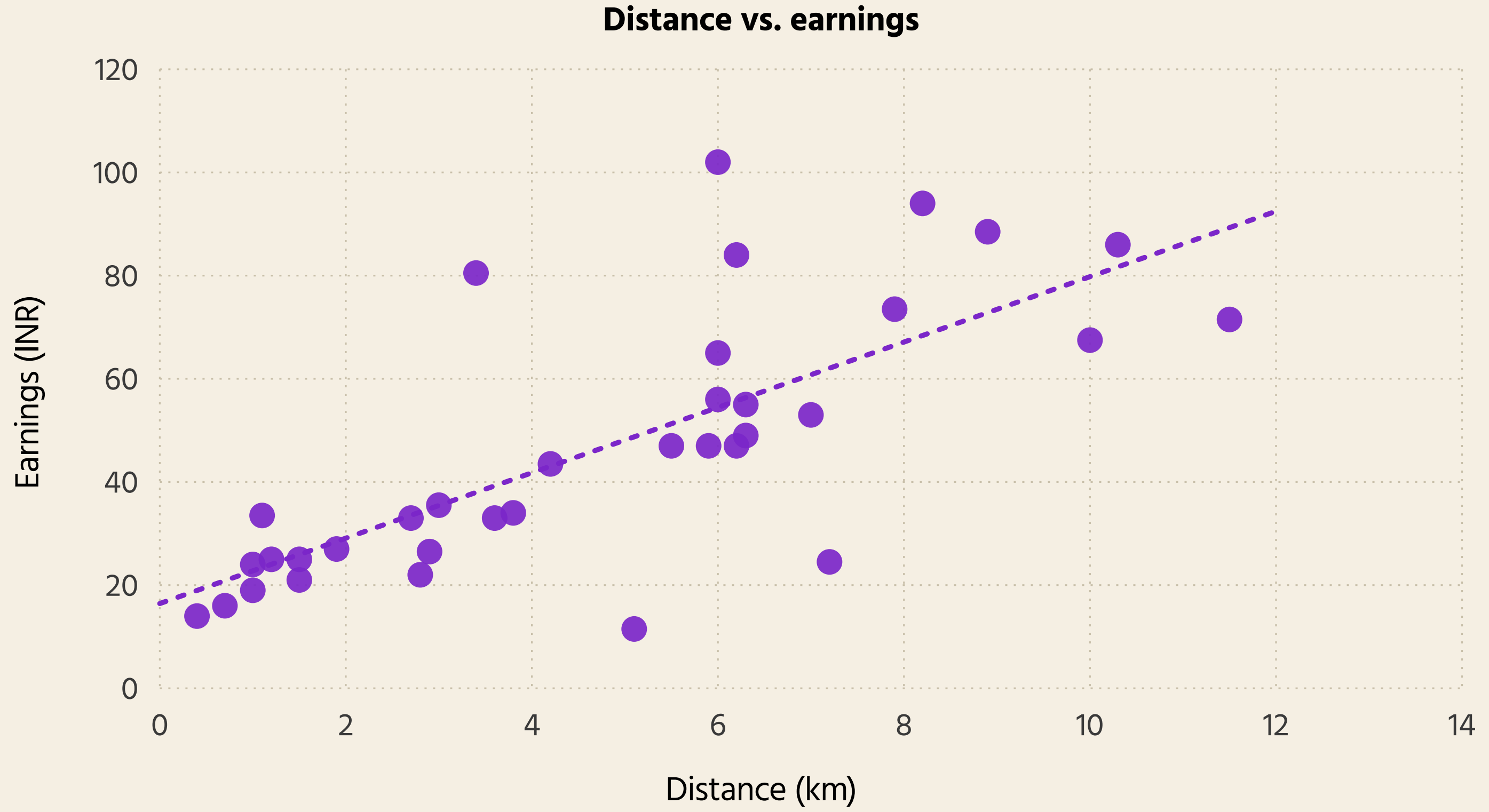

high variability, while earnings per minute varied by approximately 1.51.

**Quick commerce platforms value work in obtuse and opaque ways which cannot be determined by volume of work and effort alone.**

This can be further established by plotting earnings against distance per order (Figure 3), and time per order (Figure 4), and earnings per km. against time per order (Figure 5). As depicted below, there is a positive linear relationship between distance travelled, per order time and earnings. In the case of the former, however, data points are scattered a lot closer to the trendline when compared to per order time which has more noise despite a positive slope.

**This suggests that even though higher distance might correspond to higher earnings, workers might not get higher payouts even if they spend more time per order.**

Figure 4: Scatterplot of on-road time per order (in mins.) and earnings per order

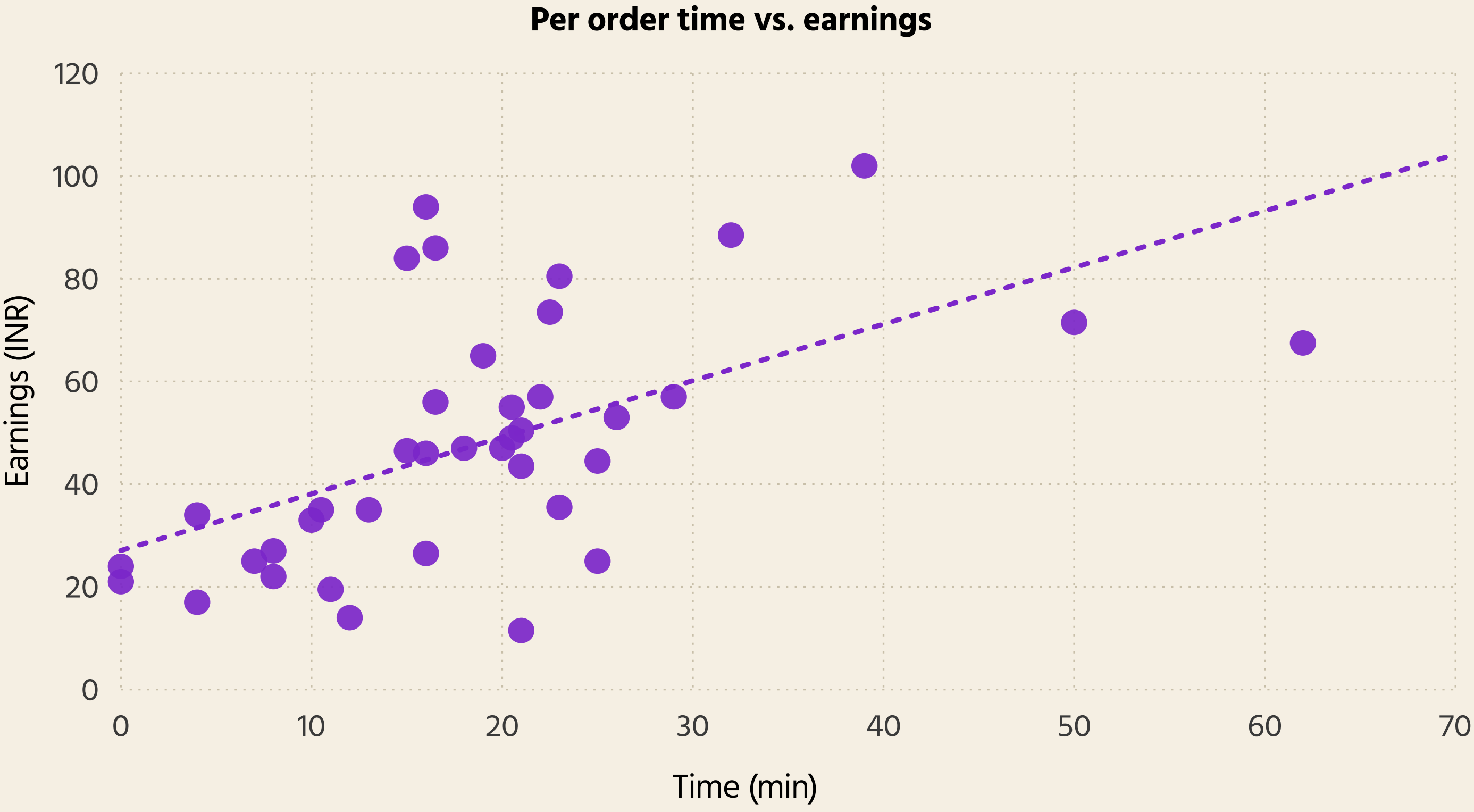

Figure 5: Scatterplot of on-road time per order and earnings per kilometer.

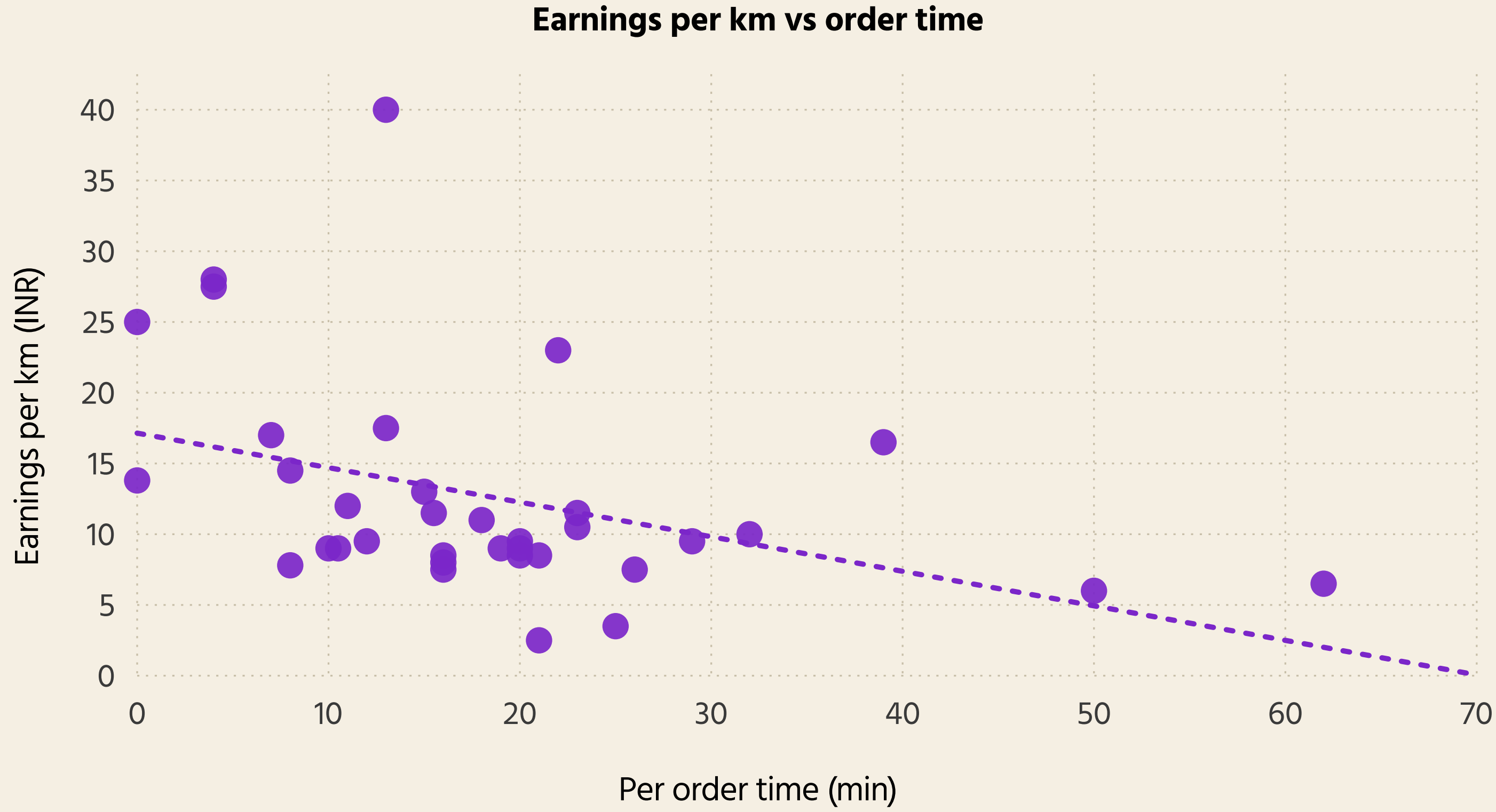

We ran a simple linear regression to test this hypothesis. We regressed earnings per order with distance travelled per order and time spent per order. Results are presented in Figure 6. Our model has a moderate goodness of fit with an R-squared value of 0.55, which indicates that approximately 55% of variations in earnings are explained by variations in distance and

## 06 SUMMARY OUTPUT: Model 1 predicts earnings per order as a function of distance travelled in kilometers and time spent in minutes per order

| Regression Statistics | |
|---|---|
| Multiple R | 0.7434799 |
| R Square | 0.55276236 |
| Adjusted R Square | 0.52922353 |
| Standard Error | 16.1266423 |
| Observations | 41 |

**ANOVA**

| | df | SS | MS | F | Significance F |
|---|---|---|---|---|---|
| Regression | 2 | 12214.3852 | 6107.19 | 23.48 | 0.0000 |
| Residual | 38 | 9882.60649 | 260.07 | | |
| Total | 40 | 22096.9917 | | | |

| | Coefficients | Standard Error | t Stat | P-value | Lower 95% | Upper 95% | Lower 95.0% | Upper 95.0% |
|---|---|---|---|---|---|---|---|---|
| Intercept | 16.3189228 | 5.34607519 | 3.05 | 0.00 | 5.4964 | 27.1414862 | 5.4963594 | 27.1414862 |
| Distance (kms) | 6.41696383 | 1.38291387 | 4.64 | 0.00 | 3.6174 | 9.2165266 | 3.61740106 | 9.2165266 |
| Time (mins) | -0.0135311 | 0.31932737 | -0.04 | 0.97 | -0.6600 | 0.63291333 | -0.6599756 | 0.63291333 |

**Figure 6: Regression results for Model 1 which predicts earnings per order as a function of distance travelled and on-road time**

time. This again suggests that **a significant proportion of workers' earnings is determined by invisible factors that are never reported to them and are difficult to compute based on the information they have.**

As reported in Figure 6, the coefficient for distance travelled per order is positive (6.42) and statistically significant at 95% confidence interval (p value is less than 0.05). This suggests a statistically significant positive relationship between earnings and long-distance orders. However, in the case of time spent per order, the coefficient is negative (-0.013) but not statistically significant (p value greater than 0.05). **This suggests that we cannot establish a linear positive relationship between earnings and time spent per order. Spending more time on an order does not necessarily translate to higher compensation.**

The time-work mismatch came out very clearly in conversations with workers. They are not compensated for time spent waiting, getting into malls and gated societies, and road barricading which are all extremely common in elite residential neighborhoods. A worker in the Zamrudpur store talked about how gated societies would lock their gates at night and block access to main roads, making workers take much longer alternate routes. Their payout did not adjust for such changes even though the distance travelled was recorded on the app.

Compounded over days, weeks, and months, this uncompensated pay has serious implications for workers. This is but one example of unpaid time spent on orders. Several workers complained of having to climb multiple floors to deliver orders that only pay them Rs. 20 in hand, while they are only being paid to arrive at the location and the physical labour of climbing is completely uncompensated.

### *II. Incentives are clearly deployed as a tool of labor control. Higher earnings per order correspond to a lower share of incentives in total earnings*

Our second dataset consisted of 23 data points around weekly earnings, including base pay, incentives, number of orders per week, and earnings per hour. Missing values under earnings per hour and no. of orders were substituted with average values for their respective ranges. Based on this we imputed total earnings, earnings per order, and share of incentives in total earnings.
On average, workers completed about 90 orders, without correcting for outliers which were several. Weekly base earnings were Rs. 2537 but with very high variability (S.D approx. 2182). Workers earned about Rs. 828 as incentives per week which again had a high variation (standard deviation = 819).

Similarly, both earnings per hour and earnings per order had high variability. And finally, and most importantly, **about a quarter of earnings (23.5%) is made up of incentives.**

Our regression results (Figure 7) from Model 2 which predicts total weekly earnings confirms the importance of incentives. The coefficient for both incentives (1.79) and no. of orders (20.29) is positive and statistically significant at 95% confidence interval, which confirms that wages are determined by gamified metrics. In fact, for every one rupee increase in incentives, our model estimates that total weekly earnings increase by Rs. 1.79. With an adjusted R-squared value of 81%, the model has moderately high goodness of fit (81% of changes in total weekly earnings can be explained with changes in the three variables – incentives, earnings per hour, and no. of orders).

This still leaves the question of what other factors impact earnings, and whether or not workers are aware of such effects at all. In our conversations, workers debated amongst themselves how their payout was determined. One worker even noted how time pressure meant that their only focus everyday was completing maximum orders, rather than inquiring about fluctuating wages.

And finally, t**he proportion of earnings made up of incentives also had high variability (S.D. of approx. 11.63) which hints at how platforms deploy incentives, bonuses, and surges as tools to regulate the size of workforce for hyperfast doorstep convenience.** The one consistent thing that we heard from workers regarding payouts was that new riders get more pay and earnings decline with time.

We then calculated correlations between orders, earnings and incentive share. There was a low yet negative correlation (-0.254) between number of orders and earnings per hour, which suggests that higher number of orders does not necessarily correspond to higher earnings.

In fact, a simple linear regression model we ran with number of orders, weekly base earnings, and weekly incentives as explanatory variables for earnings per hour reported a negative coefficient for number of orders, but this was not statistically significant at 95%

## 07 SUMMARY OUTPUT: Model 2 predicts total weekly earnings as a function of incentives, earnings per hour, and number of orders

| Regression Statistics | |
|---|---|
| Multiple R | 0.915647787 |
| R Square | 0.838410869 |
| Adjusted R Square | 0.812896796 |
| Standard Error | 1231.721523 |
| Observations | 23 |

**ANOVA**

| | df | SS | MS | F | Significance F |
|---|---|---|---|---|---|
| Regression | 3 | 149562741.1 | 49854247.04 | 32.86072197 | 1.008E-07 |
| Residual | 19 | 28825620.28 | 1517137.91 | | |
| Total | 22 | 178388361.4 | | | |

| | Coefficients | Standard Error | t Stat | P-value | Lower 95% | Upper 95% | Lower 95.0% | Upper 95.0% |
|---|---|---|---|---|---|---|---|---|
| Intercept | -133.731601 | 897.3539717 | -0.14902882 | 0.883101328 | -2011.91505 | 1744.451847 | -2011.91505 | 1744.451847 |
| Incentives | 1.797487289 | 0.478397732 | 3.757307299 | 0.001333562 | 0.796189329 | 2.798785249 | 0.796189329 | 2.798785249 |
| Earnings per hour | 1.491440576 | 6.523447932 | 0.228627651 | 0.821601059 | -12.1622929 | 15.14517401 | -12.1622929 | 15.14517401 |
| No. of orders | 20.29206925 | 6.088218359 | 3.333006153 | 0.003495355 | 7.54928178 | 33.03485673 | 7.54928178 | 33.03485673 |

**Figure 7: Regression results from Model 2 which predicts total weekly earnings as a function of incentives, earnings per hour, and no. of orders**

confidence interval.

We also found a very low positive correlation (0.029) between earnings per hour and incentives as a share of total earnings, which indicates that **time spent by workers in chasing incentive targets may not necessarily be compensated for with proportionately higher earnings.**

There also exists a low to moderate negative correlation (-0.45) between earnings per order and incentives as a share of total earnings, which suggests that **as per order earnings increase, incentives decrease and vice versa.** To confirm this trend, we ran another regression (Figure 8) that predicts incentive share in total earnings as a function of earnings per hour, number of orders, and earnings per order.

As visible from the results, the coefficient for earnings per order is negative (-0.006) and statistically significant at 95% confidence interval (p value of 0.003). This means that for every one rupee increase in earnings per order, incentive share in total earnings decreases by 0.006%, a small but significant decrease when aggregated. Higher paid orders therefore correspond to lower incentives, whereas workers are given higher incentives at lower paid orders.

During our field visits we also noticed that worker applications

## 08 SUMMARY OUTPUT: Model 3 which predicts incentive share in total earnings as a function of earnings per hour, number of orders, and earnings per order

| Regression Statistics | |
|---|---|
| Multiple R | 0.71040448 |
| R Square | 0.50467452 |
| Adjusted R Square | 0.42646524 |
| Standard Error | 0.08807245 |
| Observations | 23 |

ANOVA

| | df | SS | MS | F | Significance F |
|---|---|---|---|---|---|
| Regression | 3 | 0.15016008 | 0.05005336 | 6.45287213 | 0.00338798 |
| Residual | 19 | 0.14737838 | 0.00775676 | | |
| Total | 22 | 0.29753845 | | | |

| | Coefficients | Standard Error | t Stat | P-value | Lower 95% | Upper 95% | Lower 95.0% | Upper 95.0% |
|---|---|---|---|---|---|---|---|---|
| Intercept | 0.33567666 | 0.0767335 | 4.37457793 | 0.00032604 | 0.17507161 | 0.49628172 | 0.17507161 | 0.49628172 |
| Earnings per hour | 0.00058829 | 0.0004727 | 1.24453799 | 0.22843532 | -0.0004011 | 0.00157767 | -0.0004011 | 0.00157767 |
| No. of orders | 0.00012691 | 0.00030811 | 0.41189345 | 0.68502718 | -0.000518 | 0.00077179 | -0.000518 | 0.00077179 |
| Earnings per order | -0.0055811 | 0.00137112 | -4.0704774 | 0.00065227 | -0.0084509 | -0.0027113 | -0.0084509 | -0.0027113 |

**Figure 8: Regression results from Model 3 which predicts incentives' share in total earnings as a function of earnings per hour, no. of orders, and earnings per order.**

had several dark patterns. These patterns were UI/UX design on the application that undermine transparency and user autonomy. For instance, workers are not provided the exact breakup of their earnings (Kumar & Narayanan, 2025). Platforms also do not provide the parameters used by the algorithm to arrive at decisions relating to payouts, incentives, order allocation etc. In India, **the Government has guidelines that prohibit such patterns on consumer facing applications. No such regulation exists for platform workers.**

# Discussion & conclusions

Wage fluctuation is the major issue faced by workers in this sector, and is one of the main levers through which companies 'push' and 'pull' workers into their dynamic labour pool. Put simply, they reduce wages drastically when consumer demand is low and labour availability is high, and vice versa. Wages are regulated through gamified features making it difficult for workers to keep up, and they are always 'playing' to maximise their incentives (while base pay remains a smaller percentage of pay).

Workers in the quick commerce sector have consistently highlighted abysmal payouts, erratic rate cuts, tedious working hours, and ineffective grievance redressal systems. We learned that workers regularly go on strikes when there are changes in the structure of their earnings. However, they are stuck in a loop of fluctuating wages and changing conditions to meet target incentives; even as companies deliberately make the payout structures more opaque over time.

Systematic data collected through workers — and most critically the insights of which can be fed back to them — is an indispensable tool to address the information gap. Worker-led data will allow insight into wage calculation such that they can better determine when and how long to work, as well as how to target their demands during collective action. Workers' actions are currently reactionary and fragmented, leading to immediate repression and ineffectiveness; or short-lived wins which are immediately reversed with the next cycle of incentive structures. Continuous monitoring, whether that be of individual earnings or incentive structures, will allow workers and their unions to make more strategic decisions about action-taking as well as more authoritative claims to policymakers.

Based on the data on incentives and individual order earnings we have collected, we can already see that the only real predictor of earnings is distance travelled. Time spent on the order (even leaving aside waiting or 'dead' time, which is not being calculated) does not correlate to earnings. Crucially the allocation of work (long or short distance rides, rides with or without tips etc.) is not controlled by workers, indicating that workers have very little control over how much they can earn apart from working longer hours and aiming to deliver faster. Our findings indicate that work is intensified without any guarantees of higher pay, keeping workers running on the hamster wheel to meet incentive targets.

When scaled, our project will be able to provide a clear picture of what these incentives are, the behavioural patterns towards which they push workers, and the variable pay that forces them towards longer hours and more intensity of work. Our analysis highlights this variability; and for the first time in India, aimed to provide a way to track this over time rather than in individual snapshots. In the future, we hope to experiment with tools to provide easier ways for workers to share their data and integrate with existing organising and networks. We also hope to reach out to unions in other cities to understand the potential of expanding across other sectors and geographies.